\documentclass[10pt,conference]{IEEEtran}
\IEEEoverridecommandlockouts

\usepackage{cite}
\usepackage{amsmath,amssymb,amsfonts}
\usepackage{algorithmic}
\usepackage{graphicx}
\usepackage{textcomp}
\usepackage{xcolor}
\usepackage{graphicx}
\usepackage{booktabs}
\usepackage{url}
\usepackage{tabularx}
\usepackage{colortbl}
\usepackage{array}
\usepackage{booktabs}
\usepackage{makecell}
\usepackage{threeparttable}
\usepackage{multirow}
\usepackage{tcolorbox}
\usepackage{subcaption}

\def\BibTeX{{\rm B\kern-.05em{\sc i\kern-.025em b}\kern-.08em
    3T\kern-.1667em\lower.7ex\hbox{E}\kern-.125emX}}

\begin{document}
\title{Exploring the SECURITY.md in the Dependency Chain: Preliminary Analysis of the PyPI Ecosystem}
% Understanding SECURITY.md Role in the Dependency Supply Chain

\author{%
Chayanid Termphaiboon\textsuperscript{†}, 
Raula Gaikovina Kula\textsuperscript{§}, 
Youmei Fan\textsuperscript{*},\\
Morakot Choetkiertikul\textsuperscript{†},
Chaiyong Ragkhitwetsagul\textsuperscript{†}, 
Thanwadee Sunetnanta\textsuperscript{†}, 
Kenichi Matsumoto\textsuperscript{*} \\[1ex]
\textsuperscript{†}Faculty of Information and Communication Technology, Mahidol University, Thailand \\
\textsuperscript{*}Graduate School of Science and Technology, Nara Institute of Science and Technology (NAIST), Japan \\
\textsuperscript{§}Graduate School of Information Science and Technology, The University of Osaka, Japan \\
chayanid.ter@student.mahidol.ac.th, raula-k@ist.osaka-u.ac.jp, fan.youmei.fs2@is.naist.jp, \\
\{morakot.cho, chaiyong.rag, thanwadee.sun\}@mahidol.ac.th, matumoto@is.naist.jp
}

\maketitle

\begin{abstract}

Security policies, such as SECURITY.md files, are now common in open-source projects. They help guide responsible vulnerability reporting and build trust among users and contributors. Despite their growing use, it is still unclear how these policies influence the structure and evolution of software dependencies. Software dependencies are external packages or libraries that a project relies on, and their interconnected nature affects both functionality and security. This study explores the relationship between security policies and dependency management in PyPI projects. We analyzed projects with and without a SECURITY.md file by examining their dependency trees and tracking how dependencies change over time. The analysis shows that projects with a security policy tend to rely on a broader set of direct dependencies, while overall depth and transitive dependencies remain similar. Historically, projects created after the introduction of SECURITY.md, particularly later adopters, show more frequent dependency updates. These results suggest that security policies are linked to more modular and feature-rich projects, and highlight the role of SECURITY.md in promoting proactive dependency management and reducing risks in the software supply chain.

% Security policies, including the addition of a SECURITY.md file, are becoming increasingly essential in open-source ecosystems to improve project trustworthiness. However, their relationship with software dependencies remains ambiguous. This study investigates whether PyPI projects with a security policy differ in dependency structures from those without, and further evaluates how dependencies evolve over time in projects that adopt a security policy. Using 460 projects (211 with a security policy, 249 without), we construct dependency trees and compare three metrics: maximum depth, direct dependencies, and transitive dependencies. Statistical testing shows that projects with a security policy have significantly more direct dependencies ($\mu = 5.98$ vs. $\mu = 4.25$, $p = 0.017$), while differences in transitive dependencies and maximum depth are not significant. For temporal analysis, projects with a security policy were categorized by creation era and adoption timing. Results show that projects created in the Post-SECURITY.md era (after 23 May 2019) with delayed adoption of SECURITY.md ($> 7 \text{ days after project creation}$) are more active in modifying dependencies than projects in the Pre-SECURITY.md era that adopted SECURITY.md with a delay. These findings indicate that projects with a security policy are significantly more modular and feature-rich. This underscores the importance of proactive dependency management in secure software ecosystems and highlights the role of SECURITY.md in mitigating supply chain risks.
\end{abstract}


\section{Introduction}
\label{1_introduction}

Open-source software (OSS) plays a central role in modern development, with many projects depending on external packages from ecosystems like PyPI and npm. While this reuse speeds up development, it also introduces security risks. A recent study found that fewer than half of known vulnerabilities in PyPI packages are patched promptly, leaving many projects exposed \cite{Alfadel:EMSE2023}. This underscores the need for clear vulnerability disclosure processes. Given Python’s wide adoption and the scale of PyPI, it is a relevant setting to examine how security practices relate to dependency risks \cite{StackOverflow:Survey2025:online}. To support secure practices, platforms like GitHub have promoted the use of standardized security policies. In May 2019, GitHub introduced the SECURITY.md file as a dedicated place for security disclosure information \cite{SecurityPolicyGitDocs:online}. It typically includes instructions for privately reporting vulnerabilities, often via email rather than public issue trackers. The goal is to promote trust, ensure consistent handling of vulnerabilities, and support faster response times \cite{Kancharoendee2025}. SECURITY.md is now among the top recommended practices for maintainers \cite{BestSecurityPractice:online}. However, studies show that security-related issues are often harder to resolve, usually handled by a small group of developers, and left open longer than other issues. This highlights the value of structured policies like SECURITY.md in streamlining reports and responses \cite{Bühlmann2022}. Projects with SECURITY.md also tend to score higher in tools like the OpenSSF Scorecard, especially in areas such as branch protection and dependency update tools, suggesting a link between policy adoption and stronger project practices \cite{Kancharoendee2025}.

Managing dependencies is critical to maintaining open-source security. They expand the attack surface and allow vulnerabilities to spread across ecosystems, especially through transitive links. Prior studies show that many dependency vulnerabilities remain unresolved for months after disclosure, and a few maintainer accounts can become single points of failure \cite{Alfadel:EMSE2023,Zimmermann}. Developers’ response to these risks is influenced by how the information is presented; for example, visual dependency graphs have been shown to encourage more proactive updates than plain text reports \cite{Ragkhitwetsagul2024}. While both dependency management and security policies are essential for secure development, past research has treated them separately, focusing either on structural risks or on reporting practices. However, intelligent software engineering aims to bring these elements together. It seeks to support developers with systems that make informed, proactive decisions based on project-level signals. Our goal is to explore how SECURITY.md, as a form of human-defined metadata, relates to dependency behavior. By connecting security policies with how projects manage and evolve their dependencies, this study provides a foundation for building intelligent approaches to support secure and maintainable software ecosystems.

To address this gap, we examine the relationship between security policies and dependency behavior in the PyPI ecosystem. This study considers both aspects together to understand how the adoption of SECURITY.md relates to the structure and evolution of software dependencies. By linking project-level policy signals with real maintenance activity, we aim to support future work in intelligent dependency management. These research questions lay the groundwork for intelligent software engineering by exploring whether human-defined policies like SECURITY.md reflect meaningful patterns in how dependencies are organized and updated. Specifically, we investigate the following research questions:

\begin{itemize}
  \item \textbf{RQ1:} How do the dependency trees differ between projects with and without a security policy?
  \item \textbf{RQ2:} Do projects with a security policy have more dependencies over time?
\end{itemize}

By answering these questions, our study makes the following contributions:

\begin{itemize}
  \item We examine how the presence and timing of \texttt{SECURITY.md} adoption relate to the structure and maintenance of dependencies in open-source Python projects.
  \item We find that projects with a \texttt{SECURITY.md} file tend to have more direct dependencies, and those adopting the policy later exhibit higher rates of dependency-related commits.
  \item We observe that over 80\% of dependency modifications occur in \texttt{setup.py} and \texttt{pyproject.toml}, highlighting their central role in managing and evolving dependencies.
\end{itemize}

These findings shed light on how security policies affect dependency management and support future improvements in secure, maintainable open-source software. The dataset and analysis scripts are publicly available.\footnote{\url{https://github.com/morakotch/security-md-deps}}

 \section{Background and Related Work}
\label{2_background}
% This section provides the essential background information. We initially explain the importance of security policies and their role in open-source ecosystems, including prior studies on their adoption and effectiveness. We then discuss the impact of software dependencies on open-source projects and highlight how our research differs from existing work.

\subsection{Security Policies in Open-Source Ecosystems}
% \morakot{@Ching: please develop the following structure for the Background section}

% \morakot{start by introducing SECURITY.md — its purpose, typical content (e.g., how to report vulnerabilities), and why platforms like GitHub encourage its adoption. Include references if possible.}

% \morakot{summarize prior work on security policy adoption, especially Bee’s study. Briefly highlight its key findings related to PyPI projects, such as the prevalence of SECURITY.md, and observed differences in dependency metrics.}

Security policies guide users and contributors on how to report security vulnerabilities. On GitHub, these are typically documented in a SECURITY.md file, placed in the root, docs, or .github folder. GitHub introduced SECURITY.md\footnote{\url{https://docs.github.com/en/code-security/getting-started/adding-a-security-policy-to-your-repository}}  to promote responsible vulnerability disclosure \cite{SecurityMDIntroduction:online, SecurityPolicyGitDocs:online}. Security policies help open-source projects handle vulnerabilities in a clear and consistent way. They provide reporting instructions that build trust and ensure issues are addressed through private, well-defined channels \cite{SecurityPolicyGitDocs:online}. They also increase accountability by making maintainers’ responsibilities visible and improve response times by guiding reports to the right place, reducing the risk of public exposure. A typical SECURITY.md file specifies how to report vulnerabilities, with email being the most commonly preferred method among maintainers \cite{Kancharoendee2025}. According to Kancharoendee et al.~\cite{Kancharoendee2025}, the content of SECURITY.md files can be grouped into categories such as reporting channels, response timelines, disclosure preferences, and scope of coverage, each reflecting how project maintainers define their security practices.

% Security policies provide structured guidance on how users and contributors should report security vulnerabilities to project maintainers. On GitHub, such policies are typically documented in a SECURITY.md file, which may be placed in the repository’s root directory, the docs folder, or the .github folder. GitHub officially introduced support for SECURITY.md on 23 May 2019 to encourage responsible vulnerability disclosure \cite{SecurityMDIntroduction:online}\cite{SecurityPolicyGitDocs:online}.

% Security policies play a critical role in shaping how open-source communities address vulnerabilities. By providing clear instructions for reporting issues, they foster trust and transparency, as users and contributors know that security problems will be acknowledged and managed through a predictable process rather than being overlooked or disclosed irresponsibly \cite{SecurityPolicyGitDocs:online}. They also reinforce accountability, since maintainers publicly commit to handling vulnerabilities in a consistent manner, making their security posture visible to the community. Additionally, security policies enhance incident response by ensuring that vulnerability reports are directed through the correct channels, allowing maintainers to respond more quickly and reducing the likelihood of uncontrolled public disclosure that could otherwise benefit attackers.

% A typical SECURITY.md document specifies the channels and instructions for reporting discovered vulnerabilities, with email being the most commonly preferred method among maintainers \cite{Kancharoendee2025}.

\subsection{Software Dependencies Supply Chain in Open-Source Projects}

% \morakot{introduce the concept of software dependency chains. Define terms like direct dependencies, transitive dependencies, and depth. You can use simple examples to help clarify these concepts.}

% \morakot{cite studies that have analyzed risks or structure of dependency chains in open-source projects (e.g., studies on dependency freshness, supply chain risks, or cascading vulnerabilities).}

% \morakot{discuss the research gap — e.g., most existing works study SECURITY.md or dependencies separately, but little is known about how the presence of a security policy relates to the structure or nature of a project’s dependency chain.}

% \morakot{finish with a transition to your study — motivate why this intersection matters and how your work will begin to explore it.}

\begin{figure}[!t]
    \centering
    \includegraphics[width=\linewidth]{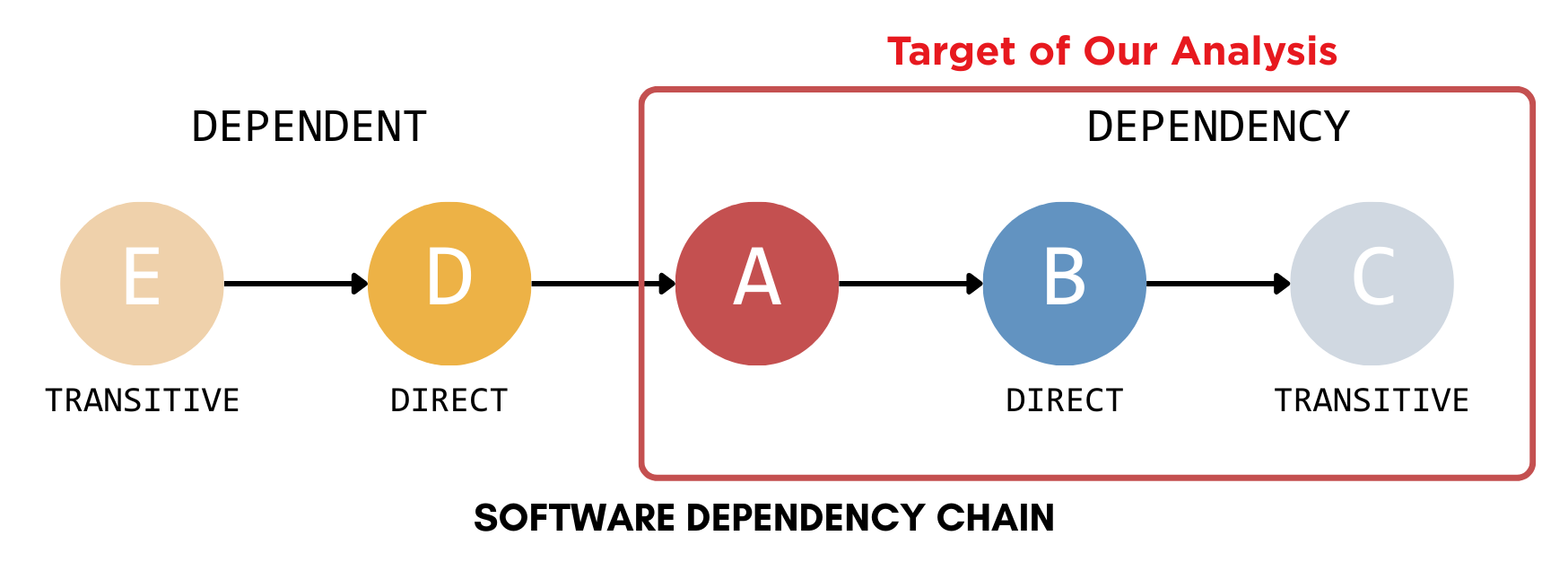}
    \caption{An example of a software dependency supply chain, showing both dependencies and dependents}
    \label{fig:swChain}
    \vspace{-.3cm}
\end{figure}

The software dependency supply chain refers to the network of software packages where projects rely on others and are also relied upon. As shown in Figure~\ref{fig:swChain}, project A is the center of our analysis. On the dependency side, if project A uses library B, B is a direct dependency of A. If B depends on library C, then C is a transitive dependency of A. Dependency depth measures the longest path from the focal project to any transitive dependency. On the dependent side, if project D uses project A, D is a direct dependent. If project E uses D, E becomes a transitive dependent of A. These relationships form a chain through which both functionality and vulnerabilities can spread. In this study, we examine only the dependency side. We analyze direct dependencies, transitive dependencies, and dependency depth to explore how the presence and timing of a \texttt{SECURITY.md} file relate to dependency structure and maintenance.

\subsection{Related Works}

Previous studies have examined security policies in open-source ecosystems, focusing on adoption and reporting practices. For example, Kancharoendee et al.~\cite{Kancharoendee2025} analyzed 679 PyPI projects, comparing declared reporting mechanisms with actual developer behavior and assessing quality using the OpenSSF Scorecard. Most projects favored private disclosure channels, such as email or GitHub advisories, although contributors often reported issues publicly. Projects with a security policy scored higher in branch protection, dependency update tools, and maintenance practices, highlighting a link between policy adoption and stronger security posture. However, the study did not examine dependency structures, leaving open the question of whether such policies relate to differences in dependency composition or evolution.

Regarding software dependencies in open-source software (OSS), several studies have highlighted their importance to project security. Prana et al.~\cite{Prana2021} showed that vulnerable dependencies can propagate risks across ecosystems. Their analysis of 450 Java, Python, and Ruby projects revealed that most vulnerabilities persist for long periods, with common types including Denial of Service and Information Disclosure. Even when fixed, vulnerabilities often take 3–5 months to resolve. Zimmermann et al.~\cite{Zimmermann} further showed that many npm projects depend on vulnerable packages, that risks spread quickly, and that a few maintainer accounts act as single points of failure. Unmaintained packages were also found to cause vulnerabilities to persist for years.

\begin{figure*}
    \centering
    \includegraphics[width=\linewidth]{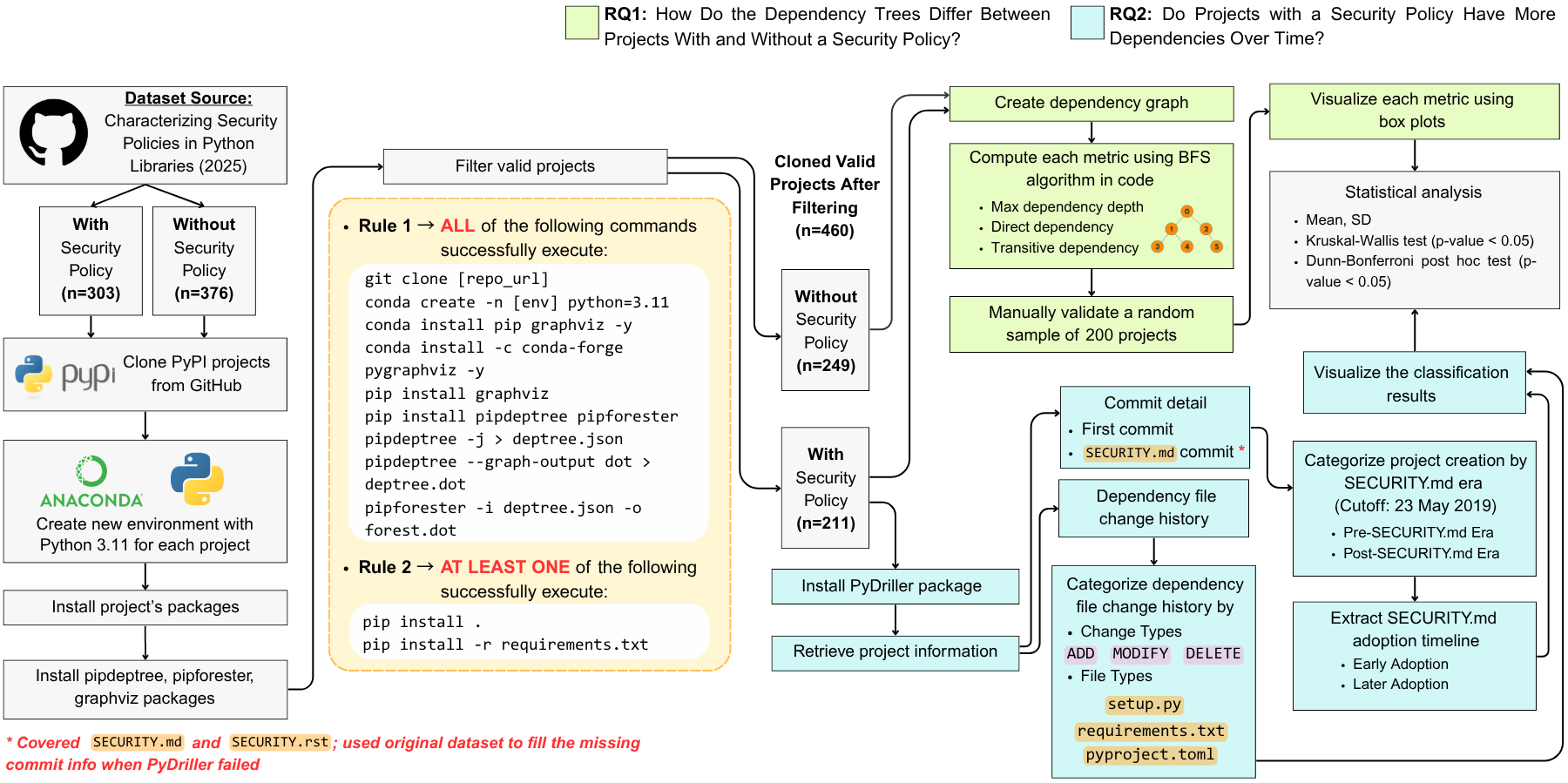}
    \caption{Overview of the research methodology}
    \label{fig:researchMethod}
    \vspace{-.3cm}
\end{figure*}

To help manage these risks, visualization-based tools have been introduced to improve developers’ understanding of dependency structures. V-Achilles, for example, uses interactive graphs to visualize transitive vulnerabilities in npm projects~\cite{Jarukitpipat2022}. It highlights both direct and transitive risks but is limited to four levels of depth and may struggle with complex graphs. A follow-up study~\cite{Ragkhitwetsagul2024} showed that 70\% of developers who used graph visualizations reprioritized updates compared to those using text-based tools. Although security policies and dependency risks have been studied, they have typically been addressed separately. Prior work has not explored whether security policies are associated with changes in the structure or evolution of a project’s dependency supply chain.

\section{Methodology}
\label{3_methodology}
% In this section, we present our methodology, describing the research process from data collection to the approach used to answer research questions.

\subsection{Overview}

Figure~\ref{fig:researchMethod} presents an overview of the research methodology. This study examines how dependency structures and their evolution differ between PyPI projects with and without a security policy. The dataset is derived from a prior study \cite{Kancharoendee2025} and categorized based on the presence of a SECURITY.md file. Dependency chains are extracted and visualized using \texttt{pipdeptree}, \texttt{pipforester}, and \texttt{Graphviz}. Projects that do not meet installation and setup criteria are excluded—92 projects with a security policy and 127 without—resulting in a final dataset of 460 projects, with 211 having a policy and 249 without. To validate extraction accuracy, a subset of projects is manually reviewed. Dependency trees are analyzed using a breadth-first search algorithm, and project histories are examined with \texttt{PyDriller} to identify changes in dependency files over time. Projects are further categorized by creation date and by the period during which a security policy was adopted. All metrics are visualized, and statistical tests are applied to assess differences between the two groups. The following sections describe each step of the methodology in detail.

\subsection{Data Collection}

This study uses a dataset from prior work \cite{Kancharoendee2025}, which includes 679 PyPI projects divided into two groups: 303 projects with a security policy and 376 without. 
%The dataset, publicly available on GitHub, also includes metadata such as the number of commits, project age, creation date, security policy creation date, and OpenSSF Scorecard results.
From the provided list of projects, each repository was cloned and executed within an isolated Python environment to extract dependency information. Dependencies were installed based on the project’s specification files. A dependency chain in this study refers to the sequence of required packages beginning with the main dependencies declared in the selected files. This study focuses on three primary dependency files: \texttt{setup.py}, \texttt{pyproject.toml}, and \texttt{requirements.txt}, as they reflect core production dependencies. Supporting tools such as \texttt{pipdeptree}, \texttt{pipforester}, and \texttt{Graphviz} were installed to generate and analyze dependency graphs. All installations were conducted using default commands. Projects that required additional or custom installation steps were excluded from the analysis. Development and auxiliary dependencies, such as those listed in \texttt{requirements-dev.txt} or \texttt{requirements-tool.txt}, were also excluded.

As shown in Figure~\ref{fig:researchMethod}, the specific steps for data collection and filtering are illustrated. A project was retained only if it passed both Rule 1 and Rule 2. Rule 1 checks whether all setup commands run successfully. These commands include cloning the repository, creating a Python 3.11 environment, and installing required tools such as \texttt{pip}, \texttt{graphviz}, \texttt{pygraphviz}, \texttt{pipdeptree}, and \texttt{pipforester}. The commands must also generate dependency graph files using \texttt{pipdeptree} and \texttt{pipforester} in DOT and JSON formats. Rule 2 checks whether at least one of the dependency installation commands succeeds. These are either \texttt{pip install .} or \texttt{pip install -r requirements.txt}. Projects that failed to satisfy both rules were excluded from further analysis. After applying these filtering rules, the final dataset consisted of 460 projects: 211 with a security policy and 249 without.

\subsection{Data Pre-Processing and Data Extraction}

This section describes how the dependency data and project metadata were processed to answer RQ1 and RQ2. For RQ1, which investigates whether dependency structures differ between projects with and without a security policy, we analyzed three key metrics: (1) maximum dependency depth, (2) number of direct dependencies, and (3) number of transitive dependencies. These metrics reflect the structural complexity and potential exposure surface of a project’s dependency supply chain. To compute these metrics, we used \texttt{pipdeptree} to extract dependency trees in JSON format, and \texttt{pipforester} to clean and structure the output. For manual validation and visualization, the structured data were converted to DOT format using \texttt{Graphviz}. A Breadth-First Search (BFS) algorithm was implemented to traverse each tree and count dependencies at each level. To validate the accuracy of the BFS implementation, we manually inspected a random sample of 200 projects.

For RQ2, which explores how dependencies evolve in projects with a security policy, we collected commit history data from the 211 valid projects using \texttt{PyDriller}. This included timestamps of the first commit, the commit introducing the security policy file, and all commits involving changes to the three primary dependency files: \texttt{setup.py}, \texttt{pyproject.toml}, and \texttt{requirements.txt}. These files were selected because they represent production-level dependencies and are the most commonly used in Python packaging. Changes were categorized as ADD, MODIFY, or DELETE to quantify the level of ongoing maintenance and dependency updates throughout the project lifecycle.

Projects were grouped based on their creation date and the timing of security policy adoption. We used 23 May 2019, when GitHub introduced support for SECURITY.md files, as the cut-off to distinguish projects created before and after this change. Each group was further divided into early and later adopters, depending on whether the policy was added within seven days of the first commit. The seven-day threshold helps identify projects that included the policy as part of their initial setup versus those that integrated it during later stages. This classification enables analysis of how adoption timing may relate to changes in dependency management. While this threshold offers a practical distinction, we acknowledge that more fine-grained analysis is possible and leave it for future work. The classification criteria are summarized in Table\ref{tab:classification}, and the analysis results are presented in the following sections.

% Projects were classified into four groups based on creation date and timing of security policy adoption. The cut-off date of 23May 2019 marks when GitHub officially introduced support for SECURITY.md files. Projects created before this date are categorized as “Pre‑SECURITY.md”, and those created after as “Post‑SECURITY.md”. Within each era, projects are further grouped as “Early Adoption” or “Later Adoption” depending on whether the security policy was added within seven days of the first commit. We selected the seven-day threshold to distinguish projects that adopted the policy as part of the initial setup from those that added it later during development. This threshold provides a clear, practical separation aligned with early project activity. We note that more fine-grained temporal analysis, such as using median adoption delay or analyzing policy updates over time, is considered for future work. Table\ref{tab:classification} summarizes these classification criteria. This classification allows us to examine how the timing and context of security policy adoption relate to the evolution of dependency management practices. A detailed analysis of the results is presented in the following sections.

\begin{table}
\renewcommand{\arraystretch}{1.25}
\centering
\caption{Project classification criteria and definitions}
\begin{tabularx}{\columnwidth}{>{\raggedright\arraybackslash}p{0.28\columnwidth} >{\raggedright\arraybackslash}X}
\toprule
\textbf{Classification Category} & \textbf{Criteria and Definition} \\
\midrule
Pre-SECURITY.md & Projects created \textbf{before 23 May 2019}, the date GitHub officially introduced support for SECURITY.md. \\
Post-SECURITY.md & Projects created \textbf{on or after 23 May 2019}, meaning they started after SECURITY.md was introduced. \\
Early Adoption & Projects that added a SECURITY.md file \textbf{within 7 days} of their first commit. \\
Later Adoption & Projects that added a SECURITY.md file \textbf{more than 7 days} after their first commit. \\
\bottomrule
\end{tabularx}
\label{tab:classification}
\end{table}

\subsection{Data Analysis}

To address RQ1, we examined structural differences in dependency trees between projects with and without a security policy. We compared three metrics: maximum dependency depth, number of direct dependencies, and number of transitive dependencies. The Kruskal–Wallis test \cite{Kruskal1952} was used to assess overall group differences. When significant differences were observed, the Dunn–Bonferroni post hoc test \cite{Dunn1964} was applied for pairwise comparisons. These nonparametric tests were chosen because the distributions of the metrics were not normal and the group sizes were unequal.

To address RQ2, we analyzed the commit history of projects with a security policy, focusing on changes to dependency specification files to understand how dependencies evolve over time. We statistically compared the total number of commits, the number of commits involving dependency files, and the distribution of modification types (e.g., ADD, MODIFY, DELETE) between early and later adoption groups. The analysis used the mean, standard deviation, Kruskal–Wallis test \cite{Kruskal1952}, and Dunn–Bonferroni post hoc test \cite{Dunn1964} to assess differences in maintenance activity. These analyses allowed us to examine whether projects with a security policy tend to update or expand their dependencies more actively over time.

\begin{figure*}
    \centering
    \includegraphics[width=0.8\linewidth]{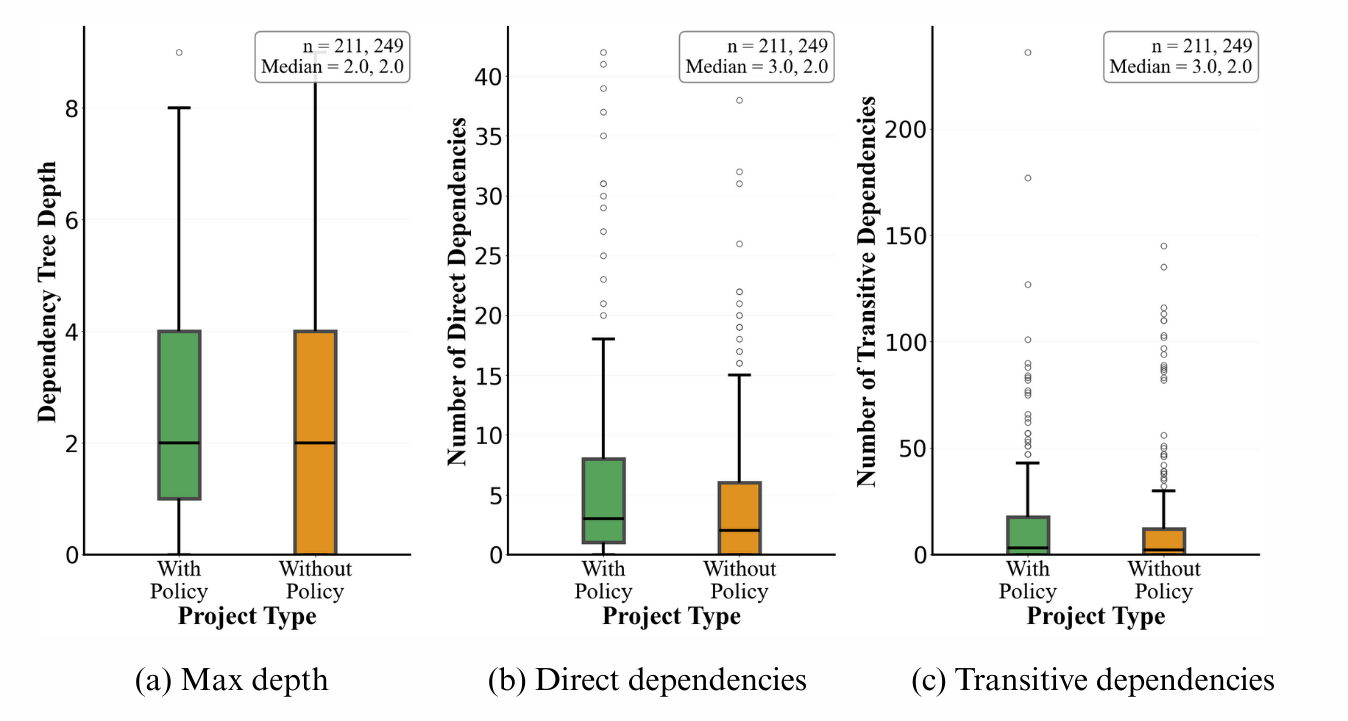}
    \caption{Distribution of dependency metrics in projects with and without a security policy}
    \label{fig:rq1}
    \vspace{-.3cm}
\end{figure*}
\section{Study Result}
\label{4_results}
This section presents the results of our study based on the two research questions.

\subsection{RQ1: How Do The Dependency Trees Differ Between Projects With And Without A Security Policy?}

Figure \ref{fig:rq1} shows the distribution of three dependency metrics for projects with and without a security policy. The maximum depth (a) is similar across both groups, with nearly identical medians and overlapping ranges. For transitive dependencies (c), projects with a policy have slightly higher values, but the distributions are wide and similar. The clearest difference appears in direct dependencies (b), where projects with a policy have a higher median and a broader upper range. This suggests that these projects tend to include more direct packages than those without a policy.

Table~\ref{tab:dependency_metrics} compares dependency metrics between 211 PyPI projects with a security policy and 249 without. Among the three metrics—maximum depth, total direct dependencies, and total transitive dependencies—only the total number of direct dependencies shows a statistically significant difference ($p = 0.017$), based on the Kruskal–Wallis and Dunn–Bonferroni tests. Projects with a security policy have a higher mean number of direct dependencies (5.98) than those without (4.25), suggesting that these projects may be more modular or incorporate more external packages. Although the mean number of transitive dependencies is also higher in projects with a security policy (15.30 vs. 13.43), the difference is not statistically significant ($p = 0.356$), likely due to high variability across projects, as reflected in the large standard deviations.

\begin{tcolorbox}[colback=gray!5!white,colframe=black,title=Answer to RQ1]
Projects with a security policy exhibit a significantly higher number of direct dependencies than those without. This difference suggests that these projects are more modular or make greater use of external packages. However, no significant differences are found in transitive dependencies or maximum depth. These findings imply that structural differences are mainly at the direct dependency level, possibly reflecting more deliberate dependency management practices in projects that adopt a security policy.
\end{tcolorbox}

\begin{table}[htbp]
\centering
\caption{Statistical comparison of dependency metrics between projects with and without a security policy}
\label{tab:dependency_metrics}
\renewcommand{\arraystretch}{1.2}
\begin{threeparttable}
\setlength{\tabcolsep}{4pt}
\begin{tabular}{lccccc}
\toprule
\textbf{Metric} &
\multicolumn{2}{c}{\textbf{With Policy}} &
\multicolumn{2}{c}{\textbf{Without}} &
\textbf{$p$-value} \\
\cmidrule(lr){2-3} \cmidrule(lr){4-5} \cmidrule(lr){6-6}
& Mean & SD & Mean & SD & KW / DB \\
\midrule
Max Depth & 2.48 & 1.97 & 2.27 & 2.01 & 0.199 / -- \\
\rowcolor{green!15}
\textbf{Direct Deps} & \textbf{5.98} & \textbf{8.19} & 
\textbf{4.25} & \textbf{5.96} & \textbf{0.017 / 0.017} \\
Transitive Deps & 15.30 & 29.45 & 13.43 & 26.45 & 0.356 / -- \\
\bottomrule
\end{tabular}
\begin{tablenotes}[flushleft] 
\footnotesize
\item \textit{Note}: KW = Kruskal–Wallis; DB = Dunn–Bonferroni (reported only when $p < 0.05$).
\end{tablenotes}
\end{threeparttable}
\end{table}

\subsection{RQ2: Do Projects With a Security Policy Have More Dependencies Over Time?}

\begin{figure*}[!t]
    \centering
    \includegraphics[width=1.05\linewidth]{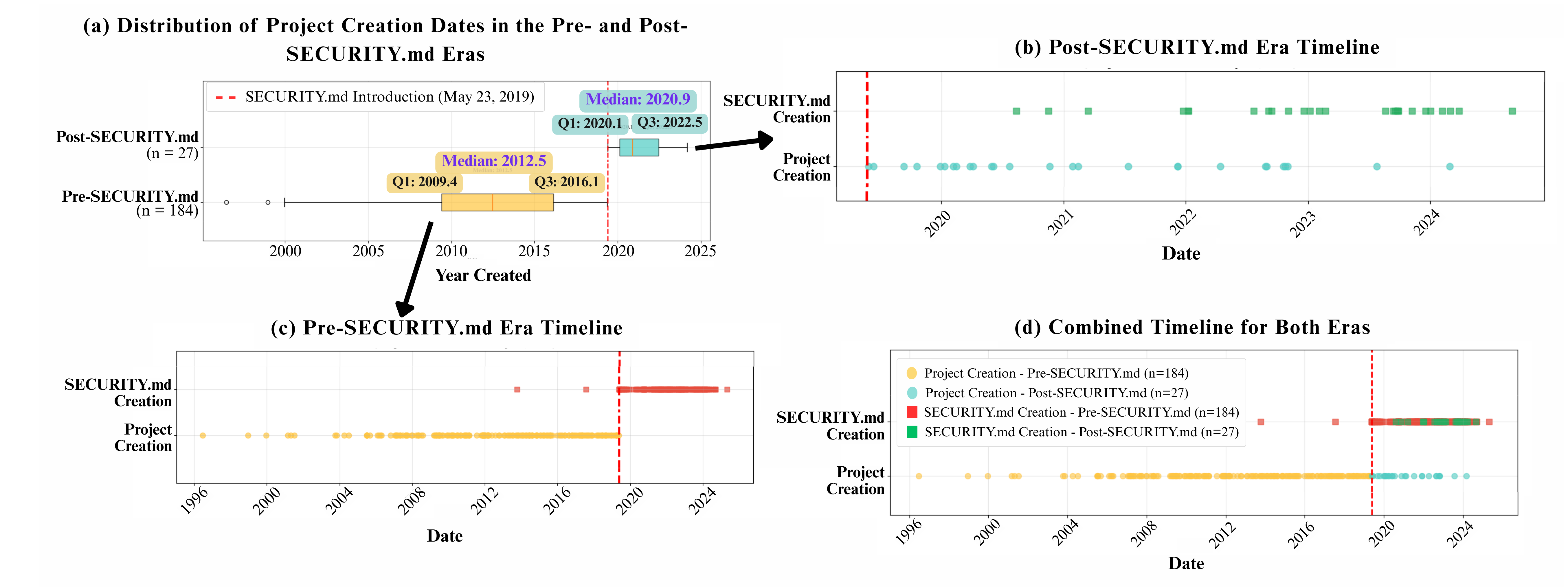}
    \caption{Distribution and timeline of project creation and SECURITY.md adoption in projects with a security policy, categorized into Pre- and Post-SECURITY.md eras}
    \label{fig:rq2_timeline}
    \vspace{-.3cm}
\end{figure*}

Figure~\ref{fig:rq2_timeline} summarizes project creation dates and SECURITY.md adoption timelines for the 211 projects with a security policy in our dataset. Projects are divided into the Pre-SECURITY.md and Post-SECURITY.md eras, using May 23, 2019—the date GitHub officially introduced SECURITY.md support—as the cutoff. The figure includes a box plot (a) and three timeline subplots (b, c, and d) to show how these projects adopted SECURITY.md over time and how adoption behavior varies between older and newer projects.

As shown in Figure~\ref{fig:rq2_timeline}(a), the majority of projects with a security policy were created before the introduction of SECURITY.md. Specifically, 184 projects were created in the Pre-SECURITY.md era. Their median creation year is 2012.5 (Q1: 2009.4, Q3: 2016.1), meaning that half of these projects were created before mid-2012 and half after. This indicates that many projects in this group have a long development history, with some dating back to the early 2000s. In contrast, only 27 projects were created after 2019, representing the Post-SECURITY.md era. These newer projects have a median creation year of 2020.9 (Q1: 2020.1, Q3: 2022.5), which is almost a decade later than the Pre-policy group. The difference of 8.4 years between the two medians shows a substantial generational gap, highlighting that security policy adoption spans across both legacy and more recent projects. The narrower interquartile range in the Post group also reflects more concentrated development activity in recent years, which may relate to more standardized development workflows and increased awareness of platform-supported security practices.

Figure~\ref{fig:rq2_timeline}(b) presents the Post-SECURITY.md projects in more detail. All 27 of these projects were created after GitHub introduced SECURITY.md support in 2019. Most of them added the SECURITY.md file relatively early in their development lifecycle, typically within the first few months after project creation. This suggests that security awareness has improved among newer projects, and that maintainers are more likely to follow GitHub’s recommended practices from the outset. The adoption timing in these projects is generally more consistent, indicating that the SECURITY.md file is now part of the initial setup rather than an afterthought. This trend may reflect a shift in community norms, where including a security policy early has become part of the standard development process.

Figure~\ref{fig:rq2_timeline}(c) focuses on the 184 Pre-SECURITY.md projects, which were created between 1996 and 2018. Most of these older projects added a SECURITY.md file only after GitHub began officially supporting it in 2019. The timing of adoption is spread out, with many projects introducing the file months or even years later. This suggests that SECURITY.md was commonly added as part of later updates or maintenance activities, rather than during initial development. The wide range of adoption times is reflected in the quartile values: the first quartile (Q1) is around 2009.4, and the third quartile (Q3) is 2016.1. This means that 50\% of Pre-SECURITY.md projects were created within that 6.7-year range, showing a long-standing base of projects that predate GitHub’s policy. These values highlight the legacy nature of many repositories and explain why the adoption of SECURITY.md was not immediate. In this study, we treat SECURITY.rst files as equivalent to SECURITY.md, as both serve the same purpose of describing the project’s security policy.

Figure~\ref{fig:rq2_timeline}(d) overlays both groups to provide a side-by-side comparison of adoption behavior. The figure clearly shows two distinct trends. Older projects, represented in the Pre group, were generally slow to adopt SECURITY.md, often adding it as part of later-stage updates. In contrast, newer projects in the Post group tend to integrate SECURITY.md shortly after their creation. This contrast reflects a broader shift in development practices: whereas older repositories treated security policies as optional or delayed concerns, newer projects increasingly treat them as essential components introduced early in the project timeline. This behavioral shift may be influenced by evolving community expectations, increased tooling support, and stronger awareness of supply chain risks in open-source software. These findings highlight the need for better integration of security policy support within modern project management tools. By embedding features that encourage or even require SECURITY.md adoption from the early stages, such tools can help maintainers meet current security expectations and reduce long-term risks.

\begin{figure}[!ht]
    \centering
    
    % Subfigure A
    \begin{subfigure}[b]{1\linewidth}
        \centering
        \includegraphics[width=0.75\linewidth]{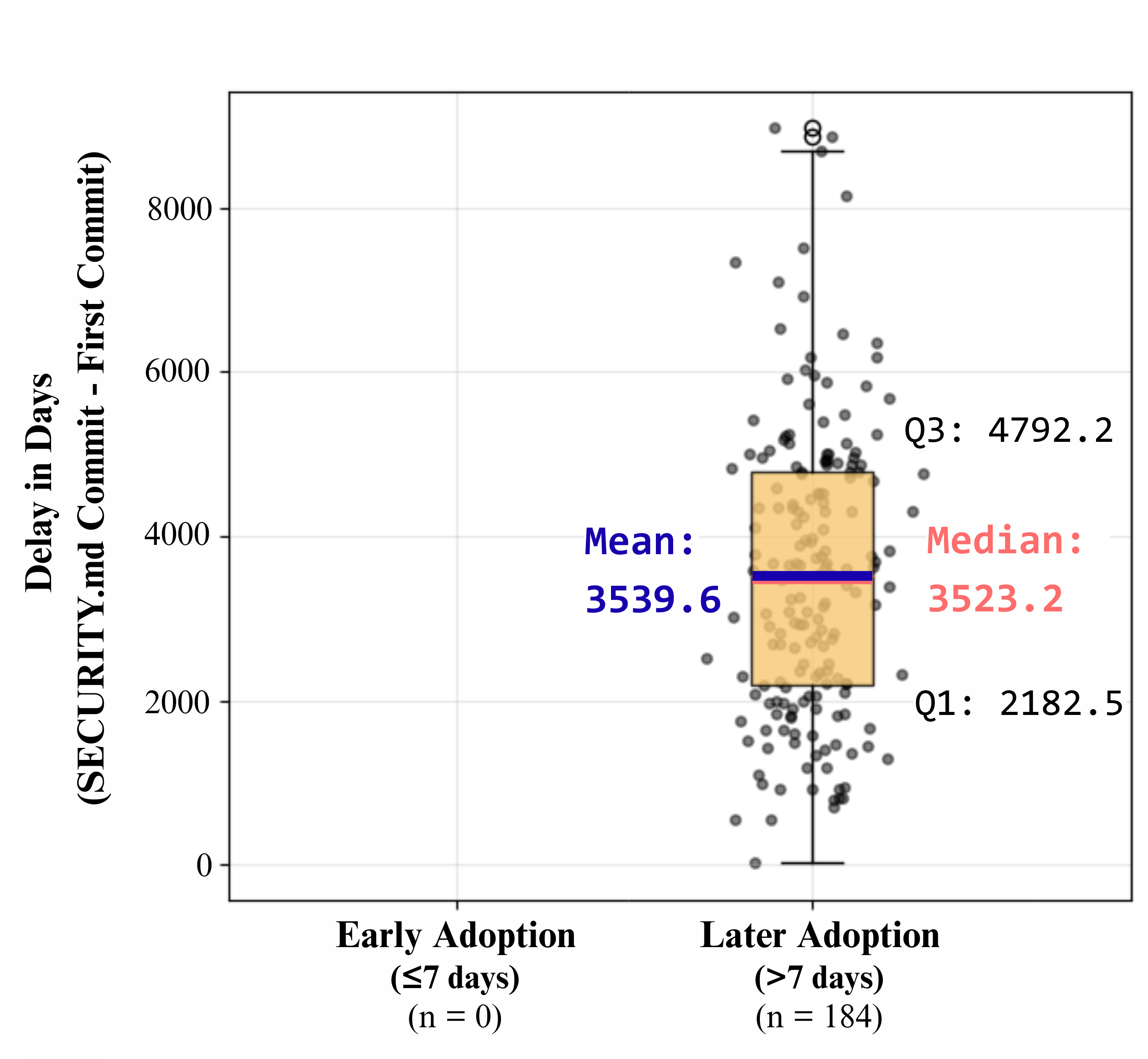}
        \caption{Pre-SECURITY.md Era Projects}
    \end{subfigure}
    
    % \vspace{0.5cm}
    
    % Subfigure B
    \begin{subfigure}[b]{1\linewidth}
        \centering
        \includegraphics[width=0.75\linewidth]{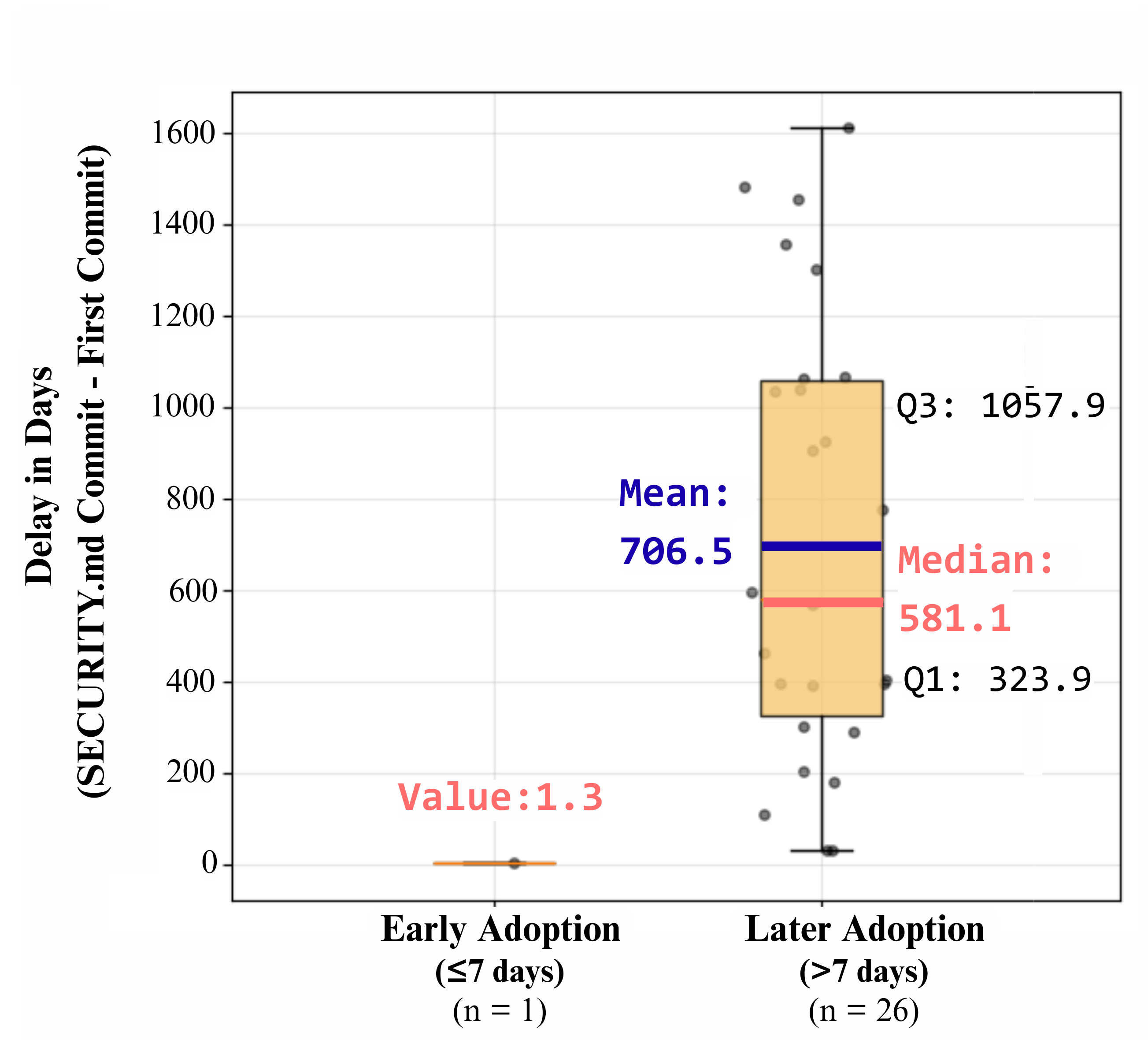}
        \caption{Post-SECURITY.md Era Projects}
    \end{subfigure}
    
    \vspace{0.3cm}
    
    \caption{Timeline of SECURITY.md Adoption Relative to Initial Commit in Pre- and Post-SECURITY.md Projects}
    \label{fig:rq2_distribution project creation date}
\end{figure}

Figure~\ref{fig:rq2_distribution project creation date} shows the delay between each project’s first commit and the creation of its SECURITY.md file, grouped by era and adoption timing. In the Pre-SECURITY.md era, all 184 projects were classified as later adopters, with a median delay of 3523.2 days (around 9.6 years). In contrast, Post-SECURITY.md era projects had a much shorter median delay of 581.1 days (about 1.6 years), showing a substantial gap of approximately 2942 days (around 8 years) between the two medians. This result indicates that, although delayed adoption remains common, newer projects tend to introduce SECURITY.md files much earlier than older ones. Note that using the first commit as the baseline may disadvantage Pre-SECURITY.md projects, so we also measured delay from May 23, 2019. Even then, these projects still adopted SECURITY.md later than Post-era ones.

Among the 27 Post-era projects, 26 (96.3\%) were still classified as later adopters, with only one project adding the SECURITY.md file within 7 days of the first commit. In comparison, no Pre-era projects qualified as early adopters. This pattern suggests that while the overall speed of adoption has improved in recent years, immediate integration of a security policy at the beginning of a project remains rare. In short, most projects took more than seven days to add a SECURITY.md file after they started.

\begin{table}
\small
\centering
\caption{Change Types Across Packaging Files in 211 Projects with Security Policies}
\label{tab:dependency_changes}
\renewcommand{\arraystretch}{1.1}
\resizebox{\columnwidth}{!}{%
\begin{tabular}{lccc@{\hskip 6pt}c}
\toprule
\textbf{Type} & \textbf{setup.py} & \textbf{req.txt} & \textbf{pyproj.toml} & \textbf{Total} \\
\midrule
MODIFY & \cellcolor{green!15}33,624 {\scriptsize(46.3\%)} & 11,406 {\scriptsize(15.7\%)} & 22,923 {\scriptsize(31.6\%)} & 67,953 {\scriptsize(93.6\%)} \\
ADD    & 989 {\scriptsize(1.4\%)} & \cellcolor{green!15}1,267 {\scriptsize(1.7\%)} & 1,178 {\scriptsize(1.6\%)} & 3,434 {\scriptsize(4.7\%)} \\
DELETE & \cellcolor{green!15}562 {\scriptsize(0.8\%)} & 450 {\scriptsize(0.6\%)} & 192 {\scriptsize(0.3\%)} & 1,204 {\scriptsize(1.7\%)} \\
\midrule
\textbf{Total} & 35,175 {\scriptsize(48.5\%)} & 13,123 {\scriptsize(18.1\%)} & 24,293 {\scriptsize(33.5\%)} & 72,591 {\scriptsize(100\%)} \\
\bottomrule
\end{tabular}
}
\end{table}

In the activities investigated in this study, Table~\ref{tab:dependency_changes} shows the distribution of change types (MODIFY, ADD, DELETE) across three packaging files in 211 projects with a security policy. Regarding the change type, MODIFY operations account for the vast majority (93.6\%) of all changes, indicating that projects typically focus on refining or updating existing dependency specifications rather than introducing new dependencies or removing outdated ones. ADD and DELETE changes are relatively limited, comprising only 4.7\% and 1.7\%, respectively. In terms of file usage, setup.py is the most actively modified file, accounting for 48.5\% of all changes. This is followed by pyproject.toml (33.5\%) and requirements.txt (18.1\%). Together, modifications in setup.py and pyproject.toml represent over 80\% of all dependency-related file changes, suggesting that these two files serve as the primary configuration points for managing dependencies in Python projects with a security policy.

\begin{table}
\footnotesize
\centering
\caption{Summary Statistics of Dependency File Changes by Group}
\label{tab:rq2_statistical_changes}
\begin{threeparttable}
\resizebox{\columnwidth}{!}{%
\begin{tabular}{@{\hskip 2pt}llc@{\hskip 4pt}ccc@{\hskip 4pt}ccc@{\hskip 4pt}cc@{\hskip 2pt}}
\toprule
\multicolumn{2}{c}{\textbf{Group}} & \textbf{Commits} &
\multicolumn{3}{c}{\textbf{Dep. Changes*}} &
\multicolumn{3}{c}{\textbf{Dep. Commits}} &
\multicolumn{2}{c}{\textbf{\% Rate** }} \\
\cmidrule(lr){1-2} \cmidrule(lr){4-6} \cmidrule(lr){7-9} \cmidrule(lr){10-11}
\textbf{Era} & \textbf{Adoption} & \textbf{Total} &
\textbf{Sum} & \textbf{Mean} & \textbf{SD} &
\textbf{Sum} & \textbf{Mean} & \textbf{SD} &
\textbf{Mean} & \textbf{SD} \\
\midrule
Pre & Early & 0 & 0 & - & - & 0 & - & - & - & - \\
Pre & Later & \textgreater1M & 52,198 & 283.7 & 407.1 & 45,300 & 246.2 & 286.6 & 8.9 & 8.1 \\
Post & Early & 53 & 36 & 36.0 & 0.0 & 36 & 36.0 & 0.0 & 67.9 & 0.0 \\
Post & Later & 101,398 & 20,357 & 783.0 & 1,873.8 & 11,220 & 431.5 & 642.4 & 16.4 & 12.7 \\
\bottomrule
\end{tabular}
}
\begin{tablenotes}[para, flushleft]
\item[*] Dependency changes include edits to \texttt{requirements.txt}, \texttt{setup.py},\\ and \texttt{pyproject.toml}.
\item[**] Percentage of commits with dependency changes; SD = standard deviation.
\end{tablenotes}
\end{threeparttable}
\end{table}

Table~\ref{tab:rq2_statistical_changes} summarizes dependency-related commit activity across project groups. The Pre-Early group had no valid samples. The Post-Early group included only one project, which showed a high dependency commit rate (67.9\%), but this is not generalizable due to the limited sample size. Among the larger groups, the Pre-Later projects (n = 184) had a lower average rate of commits involving dependency changes (8.9\%), while Post-Later projects (n = 26) showed a higher mean rate (16.4\%). This difference suggests that projects initiated after GitHub’s introduction of SECURITY.md may be more active in updating their dependency specifications.

% Table \ref{tab:rq2_statistical_changes} summarizes dependency-related commit activity across the project groups. The Pre-Early group contained no projects, while the Post-Early group (n = 1) produced an extreme dependency commit rate (67.9\%) but is not representative due to the single-project sample size. In contrast, the Pre-Later group (n = 184) exhibited a relatively low dependency commit rate (M = 8.9\%), whereas the Post-Later group (n = 26) showed a higher average rate (M = 16.4\%).

% \begin{table*}
% \centering
% \caption{Pairwise Group Comparison of Dependency Commit Rates Using Dunn-Bonferroni Test}
% \label{tab:rq2_dunnTest}
% \begin{threeparttable}
% \begin{tabular}{lllccl}
% \toprule
% \multicolumn{2}{c}{\textbf{(I) Group}} & 
% \multicolumn{2}{c}{\textbf{(II) Group}} & 
% \textbf{p-value*} & \textbf{Significant**} \\
% \cmidrule(lr){1-2} \cmidrule(lr){3-4}
% \textbf{Project Creation Era} & \textbf{SECURITY.md Adoption} & 
% \textbf{Project Creation Era} & \textbf{SECURITY.md Adoption} & & \\
% \midrule
% Pre-SECURITY.md & Later & Post-SECURITY.md & Early & 0.086647 & No \\
% \rowcolor{green!15} Pre-SECURITY.md & Later & Post-SECURITY.md & Later & 0.001048 & Yes \\
% Post-SECURITY.md & Early & Post-SECURITY.md & Later & 0.108472 & No \\
% \bottomrule
% \end{tabular}
% \begin{tablenotes}
% \item * was calculated from Dunn-Bonferroni Post Hoc Test which has $p$-value = 0.001667 ($p < 0.05$) in Kruskal-Wallis Test
% \item ** $p < 0.05$
% \end{tablenotes}
% \end{threeparttable}
% \end{table*}

\begin{table}
\footnotesize
\centering
\caption{Pairwise Group Comparison of Dependency Commit Rates Using Dunn-Bonferroni Test}
\label{tab:rq2_dunnTest}
\begin{threeparttable}
\resizebox{0.8\columnwidth}{!}{%
\begin{tabular}{lllccl}
\toprule
\multicolumn{2}{c}{\textbf{Group (I)}} & 
\multicolumn{2}{c}{\textbf{Group (II)}} & 
\textbf{p-value*} & \textbf{Sig.**} \\
\cmidrule(lr){1-2} \cmidrule(lr){3-4}
\textbf{Era} & \textbf{Adopt} & \textbf{Era} & \textbf{Adopt} & & \\
\midrule
Pre & Later & Post & Early & 0.087 & No \\
\rowcolor{green!15} Pre & Later & Post & Later & 0.001 & Yes \\
Post & Early & Post & Later & 0.108 & No \\
\bottomrule
\end{tabular}
}
\begin{tablenotes}
\item[*] Based on Dunn-Bonferroni Post Hoc ($p < 0.05$), Kruskal-Wallis $p = 0.001667$
\item[**] Significant if $p < 0.05$
\end{tablenotes}
\end{threeparttable}
\end{table}

The Kruskal–Wallis test revealed a significant difference in dependency commit rates across the project groups. Follow-up analysis using the Dunn–Bonferroni post hoc test (Table~\ref{tab:rq2_dunnTest}) showed a statistically significant difference between the Post-Later and Pre-Later groups ($p = 0.001$). This suggests that projects created after the introduction of SECURITY.md and adopting it later (after 7 days) tend to make more frequent changes to their dependency files. In contrast, no significant differences were observed between the other group pairs. These findings highlight that project creation era, rather than the timing of SECURITY.md adoption alone, may influence the frequency of dependency updates.

These findings raise a future research question: how do these dependency changes relate to broader project evolution, such as the introduction of new features, major code changes, or version upgrades? Investigating this relationship could help explain whether dependency specification updates are reactive (e.g., responding to security needs) or proactive (e.g., accompanying functional development), and how they align with software maintenance and release cycles.

\begin{tcolorbox}[colback=gray!5!white,colframe=black,title=Answer to RQ2]
Dependency activity differs significantly across groups. Projects created after the introduction of \texttt{SECURITY.md} and adopting it later (Post-Later) show significantly higher rates of dependency file changes than those created earlier (Pre-Later). Most changes occur in \texttt{setup.py} and \texttt{pyproject.toml}, which together account for over 80\% of edits. These patterns suggest that newer projects not only adopt security policies more promptly but also update their dependencies more actively, indicating stronger maintenance practices and increased attention to security.
\end{tcolorbox}

\section{Discussion and Implications}
\label{5_discussion}

Our findings provide evidence that the adoption of a SECURITY.md file in Python projects is not only a policy marker but also an indicator of a diligent, security-conscious maintenance team, reflecting real maintenance behavior rather than being a direct cause of improved practices. Specifically, projects with a SECURITY.md file tend to have more direct dependencies, suggesting greater modularity or functionality. Moreover, projects that adopt SECURITY.md later in their lifecycle show higher levels of dependency-related commits, indicating more active or frequent maintenance on dependency configurations. This pattern suggests that projects which take longer to adopt SECURITY.md may be larger or more mature, thus requiring more effort to manage dependencies. Alternatively, delayed adoption may indicate greater awareness of dependency risks and a subsequent push to improve maintenance practices. The high concentration of edits in \texttt{setup.py} and \texttt{pyproject.toml} highlights the central role of these files in Python dependency management and presents them as key targets for monitoring and tooling. These results offer several implications:

\begin{itemize}
\item For researchers, the observed link between policy adoption and dependency activity opens new opportunities to study software evolution by combining social signals (e.g., policy files) with technical data (e.g., commits).
\item For tool developers, the findings highlight \texttt{setup.py} and \texttt{pyproject.toml} as priority targets for supporting dependency tracking and automation.
\item For maintainers, adopting SECURITY.md is associated with higher maintenance activity, reinforcing its role as a practical indicator of commitment to project upkeep.
\end{itemize}
\section{Conclusion}
\label{6_conclusion}

This study examined the relationship between SECURITY.md adoption and dependency management in PyPI projects. We found that projects with a SECURITY.md file tend to have more direct dependencies, and those adopting the policy later show higher rates of dependency-related commits, especially in \texttt{setup.py} and \texttt{pyproject.toml}. These results suggest that lightweight security policies may signal stronger maintenance practices. Future empirical work could explore how adoption patterns vary across ecosystems and project characteristics, and also delve into the qualitative nature of dependency file changes. For practice, lightweight and automated tools that incorporate policy awareness could help maintainers manage dependencies more effectively, identify update needs, and sustain long-term project health.

\bibliographystyle{ieeetr}
\bibliography{bibliography}

\end{document}